\documentclass[12pt,preprint]{aastex}

\begin{document}

\title{Initial Observations of Sunspot Oscillations Excited by Solar Flare}
\author{A. G. Kosovichev}
\affil{W.W.Hansen Experimental Physics Laboratory, Stanford
University, Stanford, CA 94305, USA}
 \author{T. Sekii}
 \affil{National Astronomical Observatory of Japan,
Mitaka, Tokyo 181-8588, Japan}
%\date{\today}
%\maketitle
%\Large

\begin{abstract}
Observations of a large solar flare of December 13, 2006, using
Solar Optical Telescope (SOT) on Hinode spacecraft revealed
high-frequency oscillations excited by the flare in the sunspot
chromosphere. These oscillations are observed in the region of
strong magnetic field of the sunspot umbra, and may provide a new
diagnostic tool for probing the structure of sunspots and
understanding physical processes in solar flares.
\end{abstract}
\section{Introduction}
Solar flares represent a process of conversion of magnetic energy
into heat, kinetic energy of plasma eruptions and high-energy
particles. Solar flares may excite various types of oscillations and
waves in various layers of the Sun, from "sunquakes" in the interior
\citep{Kosovichev1998} to coronal Morton waves \citep{Moreton1960}
and coronal loop oscillations \citep{Aschwanden1999}. The mechanisms
of these oscillations and waves are not fully understood yet, but
obviously related to the energy release and transport properties.
For instance, the seismic response ("sunquakes") are believed to be
related to the hydrodynamic impact on the lower atmosphere and
photosphere by shocks generated in the area heated by high-energy
electrons. The spatial-temporal properties of the seismic wave
source are closely related to the properties of hard X-ray source
\citep{Kosovichev2006a,Kosovichev2006b}. The sunquakes represent
packets of high-frequency acoustic waves traveling through the Sun's
interior. The waves propagate through the active regions and
sunspots with strong magnetic field without significant distortion
of the wave front and large changes in the travel times. The
magnetic field effects in the sunquake waves, which to some extent
are obviously present, have not been detected.

Here, we report on observations from Hinode spacecraft of a
different type of oscillations excited by a solar flare. These
oscillations are observed in the chromosphere of the sunspot umbra
and inner penumbra, and, thus, probably represent some kind of
magnetohydrodynamic oscillatory modes. Unfortunately, the relatively
low cadence of the observations did not allow us to identify the
specific mode of the oscillations. Nevertheless, these oscillations
carry potentially interesting information about the flare energy
release and transport and properties of sunspots, and deserve
further observational and theoretical studies.

\section{HINODE Observations of December 13, 2006, Flare}

\begin{figure}
\centerline{\includegraphics[scale=0.8]{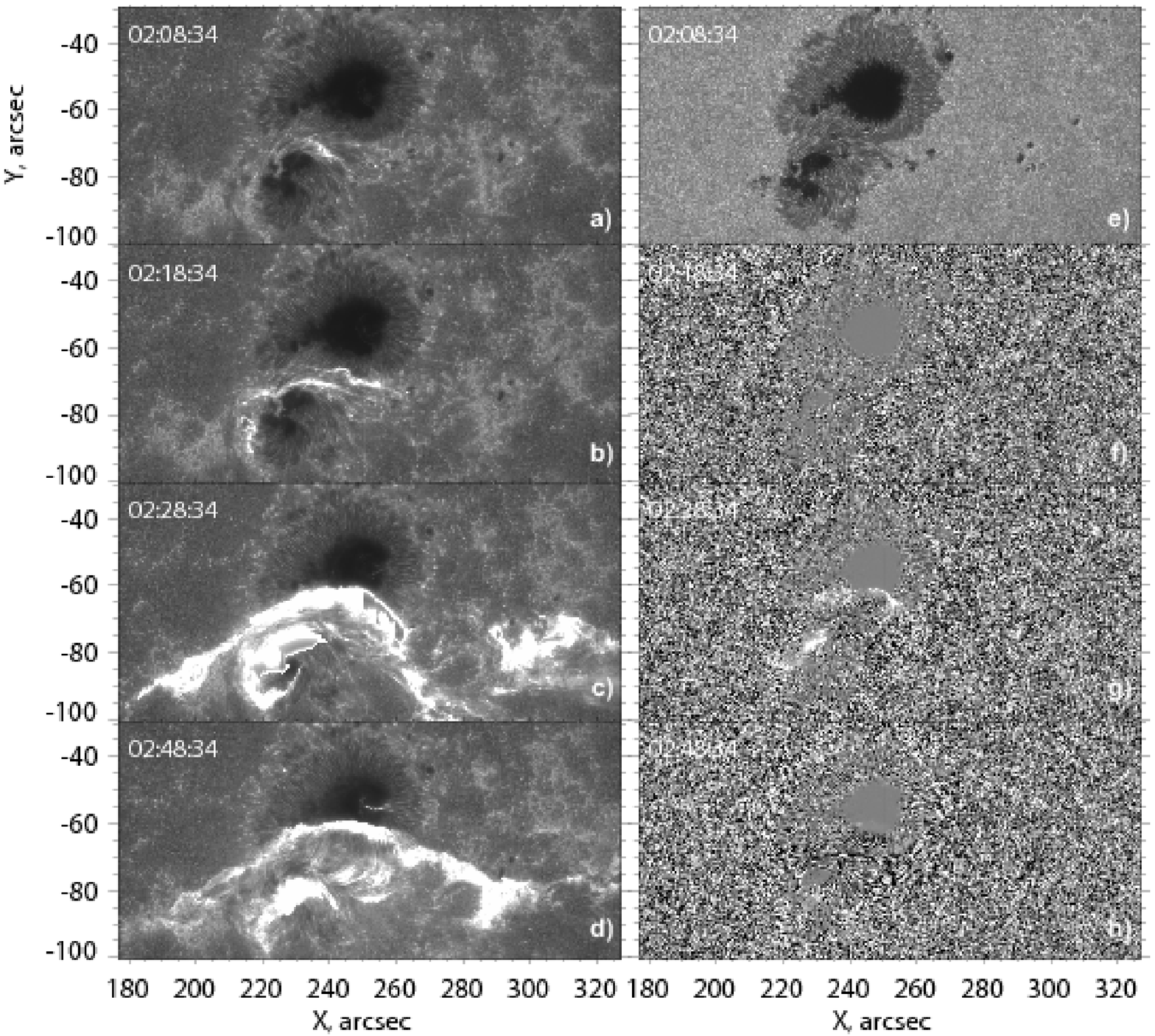}} \caption{A sample
of Hinode/SOT images of the solar flare of December 13, 2006, taken
in Ca~II~H line (a-d) and in the G-band (e-h). To reveal better the
flare signal in the G-band, the preflare image (panel e) was
subtracted in panels f-g.} \label{fig1}
\end{figure}

Images of the solar flare of December 13, 2006, were obtained by the
Solar Optical Telescope (SOT) onboard the solar space mission Hinode
in two spectral filters, the molecular line G-band and Ca~II~H. In
each filter the images were obtained every 2 minutes with the
spatial resolution of 0.2 acrsec. The pixel size of the $2048\times
1024$ images was 0.109 arcsec. The first observations of sunspot
oscillations from Hinode by \citet{Nagashima2007} have shown that
the G-band signal is suppressed in the sunspot umbra, but the
Ca~II~H data reveal high-frequency oscillations with a peak around 6
mHz, originally discovered by \citet{Beckers1969}. We also use the
Ca~II~H images in this study (Fig.~\ref{fig1}a-d. The images were
carefully tracked to remove the proper motion of the large sunspot.

A first signature of the flare appeared in the Ca~II~H images at
about 2:08 UT between two sunspots of the opposite magnetic polarity
(Fig.~\ref{fig1}a). The energy release was probably triggered by
strong shearing flows around the following (smaller in size) sunspot
\citep{Zhang2007}. Strong shearing flows have been observed also in
other flares \citep[e.g.][]{Deng2006}. The flare emission quickly
extended during the next 10 min, forming a two-ribbon structure
(Fig.~\ref{fig1}b). The ribbons expanded in length and separated
from from each other moving into the sunspot areas
(Fig.~\ref{fig1}c). The two-ribbon structure is also clearly visible
in the G-band images (Fig.~\ref{fig1}g) indicating that the
perturbation extends into the photosphere. At this stage a bright
arcade of connecting loops with the footpoints in the ribbons
appeared in the Ca~II~H images (Fig.~\ref{fig1}d). A movie of the
image differences revealed a characteristic oscillatory pattern in
the umbra of the larger sunspot after the start of the flare. The
oscillations appear to form traveling waves with a more-or-less
regular wave front.
%Figure~\ref{fig2} shows 12
%consecutive frames of this movie. The round-shaped structure in the
%middle is the sunspot umbra (best visible in the first image at
%2:18:37). It is surrounded by filamentary radial structures of
%penumbra. The bright and black structure at the bottom of the
%subsequent images is the flare ribbon moving in the sunspot umbra.
%The large-scale circular structures best visible at the umbra
%boundary in the bottom row (02:34;37-02:40:38) are the sunspot
%oscillations excited by the flare. The oscillations propagated to
%the umbra-penumbra boundary, and the signal was lost in the
%penumbra, which have much stronger background oscillations than the
%umbra.

\begin{figure}
\centerline{\includegraphics[scale=1]{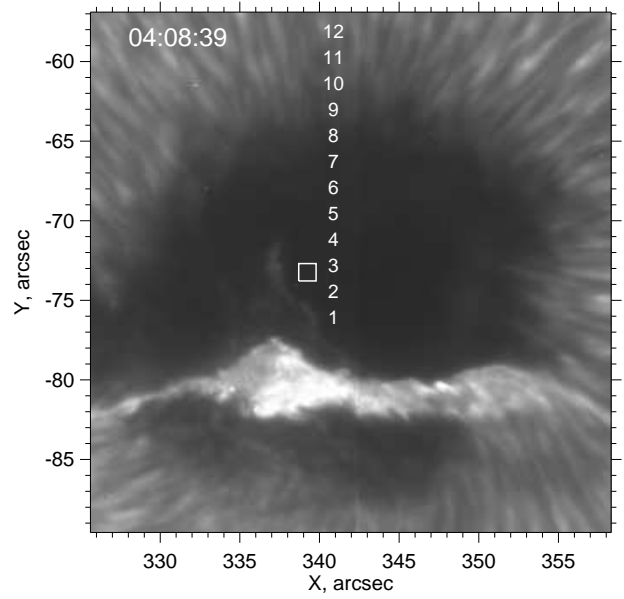}} \caption{The image of
the sunspot umbra with a portion of the flare ribbon at 04:08:39
near the end of the solar flare. The numbers from 1 to 12 show the
position of the intensity variation plots in Fig.~\ref{fig3}. The
distance between the points is approximately 1.6 Mm. The small
rectangular box is used to show long-term intensity variations in
Fig.~\ref{fig5}.} \label{fig2}
\end{figure}

\begin{figure}
\centerline{\includegraphics[scale=0.8]{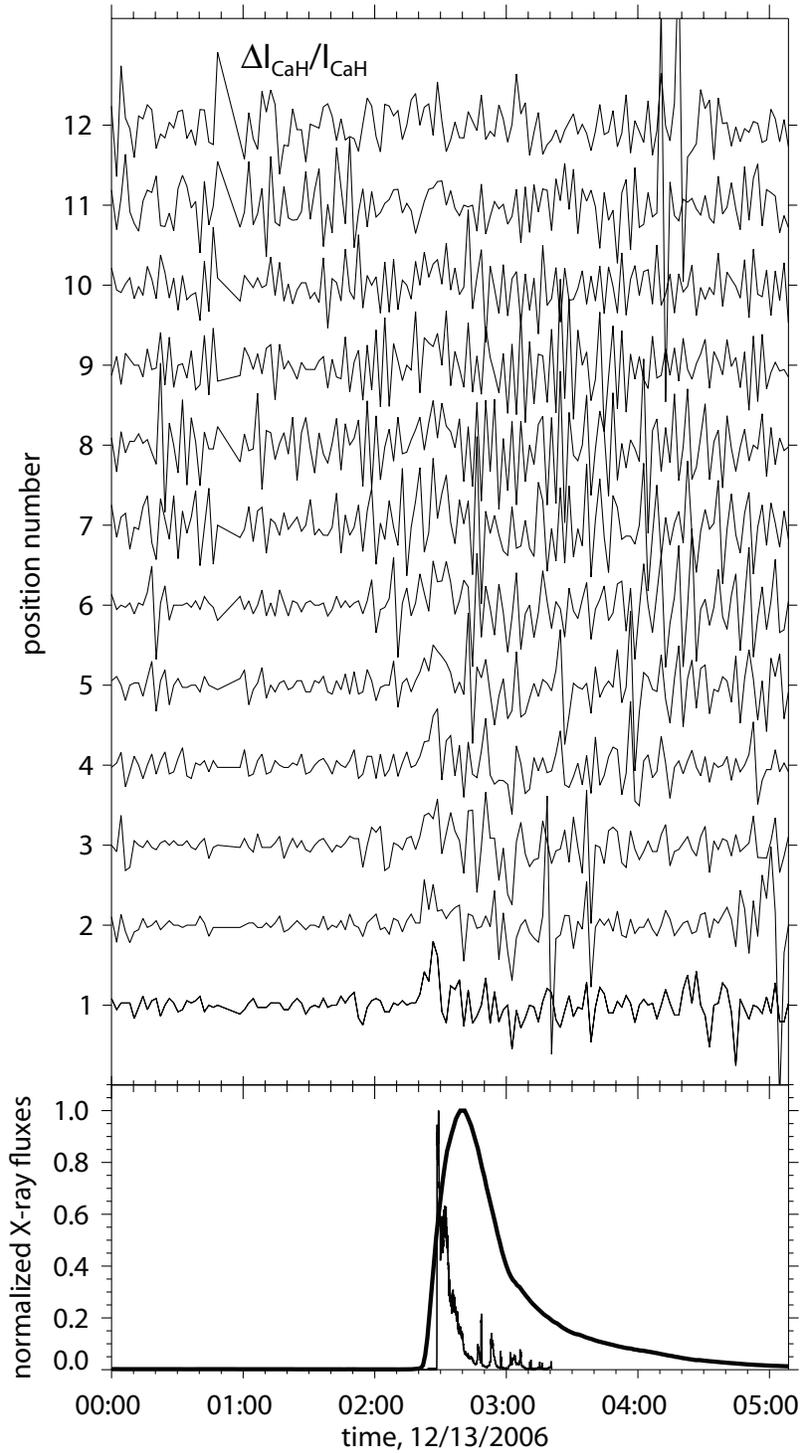}} \caption{ a) The
relative intensity variations at 12 position within the sunspot
umbra and penumbra indicated in Fig.~\ref{fig2}; b) The soft x-ray
flux measured by GOES-10 (thick smooth curve) and hard X-ray flux in
the range 25-50 keV, measured by RHESSI (thin sharp curve).
}\label{fig3}\end{figure}

\begin{figure}
\centerline{\includegraphics[scale=0.7]{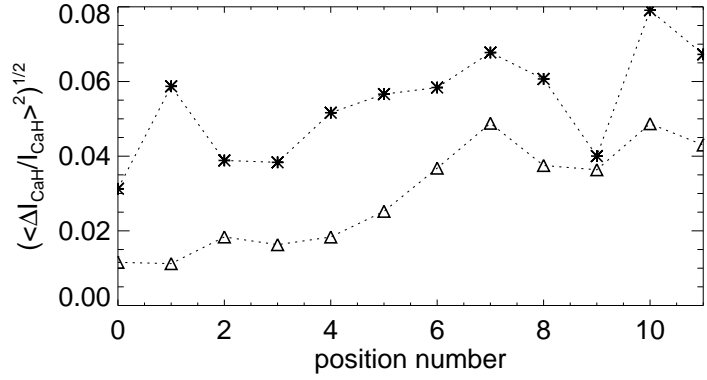}} \caption{RMS of the
relative intensity oscillation signals 2 hours before the flare,
0:00-2:00 UT, (triangles) and 2 hours after the flare, 3:00-5:00 UT
(stars).}\label{fig4}\end{figure}

To illustrate the temporal behavior of the oscillations we plot the
relative variations of the Ca~II~H intensity in 12 points (indicated
by their position numbers in Fig.~\ref{fig2}) at different distances
from the final location of the flare ribbon, in Fig.~\ref{fig3}, and
compare with the variations of the soft X-ray flux from GOES-12 and
hard X-ray in 50-100 keV energy range from RHESSI (the first few
minutes of the flare are missing in RHESSI observations, which
started at 2:28). Evidently, the oscillations in the umbra started
immediately after the hard X-ray peak and before the soft X-ray
maximum. The oscillation amplitude of the flare-excited oscillations
exceeds the amplitude of preflare oscillation in the umbra by a
factor 2-4. The RMS of the relative intensity variations
(Fig.~\ref{fig4}) shows that the relative oscillations in the
central part of the umbra changes from 0.01-0.03 to 0.04-0.06. The
relative amplitude increase is smaller in the outer umbra. The flare
oscillation signal is essentially lost in the penumbra.

It can be seen from Fig.~\ref{fig3} that the oscillations excited by
the flare have higher frequency than the oscillations prior the
flare. The characteristic period of these oscillations is probably
shorter than 3 min. Unfortunately, the 2-min cadence of the Hinode
observations did not allow us to investigate the spectral and
dispersion properties of these oscillations. Their amplitude may be
significantly underestimated because of the poor sampling.

\begin{figure}
\centerline{\includegraphics[scale=0.7]{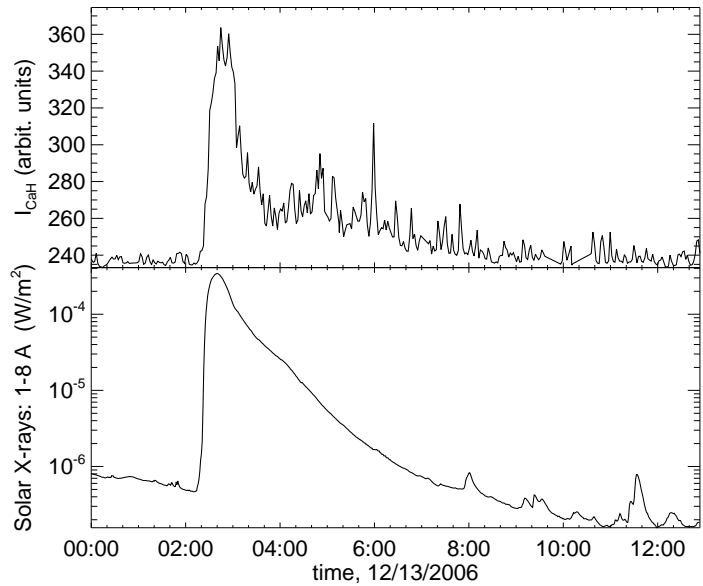}} \caption{a) Ca~II~H
intensity variations in a small umbra area shown by a rectangular
box in Fig.~\ref{fig2}. b) Soft X-ray flux from GOES-12 satellite.
}\label{fig5}\end{figure}

After the flare, the observations seem to show intrusions of hot
flare material into the umbra area in addition to the flare ribbons.
These cause variations of the Ca~II~H intensity in the umbra after
the flare, thus masking and distorting the oscillation signal. Thus,
it is rather difficult to follow the evolution of the oscillations
and estimate the lifetime. Figure~\ref{fig5} shows the Ca~II~H
intensity for longer than 12 hours in a $10\times 10$-pixel area
located in one of the least disturbed regions of the umbra. It seems
that the oscillation amplitude returned to the pre-flare level after
6 hours. However, the lifetime is not yet determined.

\begin{figure}
\centerline{\includegraphics[scale=0.7]{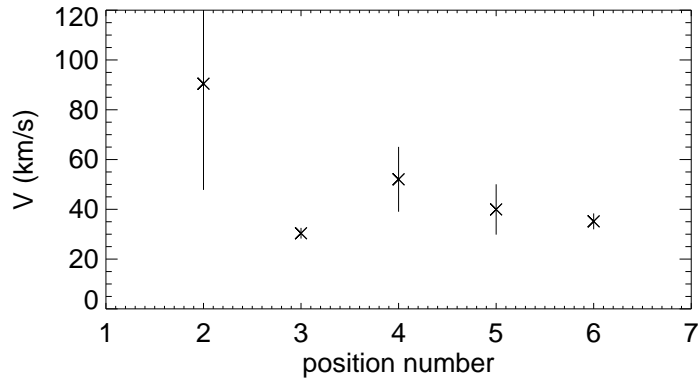}} \caption{Estimates
of the mean wave speed between 2:28 and 3:28~UT at 5 positions in
the sunspot umbra, shown in Fig.~\ref{fig2}.
}\label{fig6}\end{figure}

An interesting question is whether the observed oscillations
represent traveling or standing waves. Unfortunately, the low
cadence of this observing run on Hinode did not allow us to
convincingly answer this. Nevertheless, there is a hint that the
waves were traveling North-ward from the flare ribbon during a short
period after the flare impulse, and then the oscillations became
more random. The traveling wave feature can be seen in
Fig.~\ref{fig3} as a slight positive slope (from vertical) in the
wave signals at points 1--6 between 2:30 and 3:00. To get an
estimate of the wave speed, these signals were re-sampled to a
30-sec cadence and then cross-correlated with each other. The time
lags were measured by fitting second-degree polynomials to the
cross-correlation functions, and determining their maxima. This
procedure was repeated for 10 North-South columns along the
locations, 1--6, numbered in Fig.~\ref{fig2}, and the mean and
standard deviation values of these measurements are shown in
Fig.~\ref{fig6}. It seems that the wave speed is of the order of
50-100 km/s. This is much higher than the sound speed in the
chromosphere. Thus, these waves must be of an MHD type. In general,
the physics of sunspot oscillations is quite complicated.
Understanding of this requires numerical modeling
\citep[e.g.][]{Khomenko2006}.

\section{Discussion}

The observations of the solar flare of December 13, 2006, from
Hinode reveal a new type of flare-excited oscillations. The
oscillations observed in Ca~II~H images appeared in the
chromospheric layers of the sunspot umbra immediately after the
impulsive phase. They had the amplitude 2--4 times larger than the
pre-flare oscillations in the umbra. Also, their frequency seemed to
be higher. There is a weak evidence that during the first 30-40 min
the oscillations represent waves traveling through the umbra in the
direction away from the flare ribbon with a speed of 50--100 km/s.
Then, the oscillation become more irregular with some occasional
wave packets. The lifetime of these oscillations is probable about 8
hours. The uncertainties in the data analysis and interpretation are
caused by the low cadence of the Hinode observing program during the
flare. The images were taken every 2 min while the characteristic
oscillation period is about 3 min. Most of the oscillation power is
probably even at the shorter periods. Thus, for detailed studies of
these oscillations it will be important to increase the image
cadence.

The image cadence should be sufficiently high to capture the initial
waves traveling with a speed of $\sim 100$ km/s according to our
preliminary estimates. This speed indicates that the waves are of an
MHD type, and if their speed is of the order of magnitude of the
Alfven speed then they should propagate rather low in the sunspot
chromosphere. Thus, simultaneous observations in photospheric lines
would be interesting. The oscillation amplitude was several times
higher than the amplitude of preflare umbral oscillations, which can
be as high as  5--6 km/s \citep{Yoon1995}, and thus may reach
supersonic velocities of 10--20 km/s.

Sunspot oscillations have been studied intensively for many years
(for a review see \citet{Staude1999}) but these Hinode observations
seem to be first that show enhanced oscillations in the umbra,
associated with a solar flare. Further investigations of these
oscillations are of great interest for understanding the processes
in solar flares and sunspots.

\acknowledgements

 Hinode is a Japanese mission developed and
launched by ISAS/JAXA, with NAOJ as domestic partner and NASA and
STFC (UK) as international partners. It is operated by these
agencies in co-operation with ESA and NSC (Norway). This work was
partly carried out at the NAOJ Hinode Science Center, which is
supported by the Grant-in-Aid for Creative Scientific Research "The
Basic Study of Space Weather Prediction" from MEXT, Japan (Head
Investigator: K. Shibata), generous donations from Sun Microsystems,
and NAOJ internal funding.

\clearpage

%\begin{figure}
%\centerline{\includegraphics[scale=0.8]{f2_new.eps}} \caption{ A
%series of the image differences in the Ca~II H intensity of the
%larger sunspot (located close to the top boundary of Fig.~\ref{fig1}
%images). } \label{fig2}
%\end{figure}

\clearpage

\end{document}